# Control of Coherent Synchrotron Radiation and Micro-Bunching Effects During Transport of High Brightness Electron Beams

*D.R. Douglas, S.V. Benson, A. Hutton, G.A. Krafft, R. Li, G.R. Neil, Y. Roblin, C.D. Tennant, C.-Y. Tsai*

Thomas Jefferson National Accelerator Facility, Newport News, VA 23606

## Abstract

Beam quality preservation during transport of high-brightness electron beams is of general concern in the design of modern accelerators. Methods to manage incoherent synchrotron radiation have been in place for decades [1]; as beam brightness has improved coherent synchrotron radiation (CSR) and the microbunching instability ($\mu$BI) have emerged as performance limitations. We apply the compensation analysis of diMitri, Cornacchia, and Spampinati [2] – as previously used by Borland [3] – to the design of transport systems for use with low-emittance beams, and find that appropriately configured second order achromats [4] will suppress transverse emittance growth due to CSR and appear to limit $\mu$BI gain.

## Concept

Reference [2] gives a detailed analysis of optics compensation/suppression of CSR-driven emittance growth. It is shown therein that beam transport lines with appropriate bending symmetry and properly controlled evolution of betatron phase advance will undergo emittance compensation of CSR-induced phase space distortion. A second-order achromat [4] composed of superperiods that are individually linearly achromatic and isochronous meets all requirements stated in [2] for this suppression of CSR effects. As described by Borland [3], this is particularly evident for an achromat based on an even number of superperiods. In this case, any CSR-induced momentum shift will be paired to a matching shift at a downstream location with the same lattice parameters and the same bunch length; the transverse phase space configuration is however "inverted" by the (modulo) half-betatron wavelength phase separation. The impact of the second momentum shift therefore completely cancels (to linear order) that of the first.

This result could in principle be generalized to second order achromats of arbitrary (odd) periodicity by noting the compensation mechanism is – on a longitudinal slice-by-slice basis – identical to that of the "sum of the n complex $n^{th}$ roots of unity" suppression on which a second order achromatic is based [4].

We have applied this recipe to the design of a number of recirculation arcs for use in recirculated and energy recovering linacs and have found excellent suppression of CSR-induced growth in transverse emittance. Of considerable additional interest is also the observation that these beam line designs also manifest little or no evidence of microbunching; an initial analysis – described below – in fact indicates that this configuration has very low microbunching gain.

Notice: This manuscript has been authored by Jefferson Science Associates, LLC under Contract No. DE-AC05-06OR23177 with the U.S. Department of Energy. The United States Government retains and the publisher, by accepting the article for publication, acknowledges that the United States Government retains a non- exclusive, paid-up, irrevocable, world-wide license to publish or reproduce the published form of this manuscript, or allow others to do so, for United States Government purposes.

**Example 1: Low Energy Recirculation Arc**

A first example of the emittance compensation process under consideration is provided by a three-bend-achromat (TBA)-based arc intended for use in compact, high-current energy recovered linacs (ERLs), such as the drivers for electron cooling systems in electron-ion colliders. The structure of a single reflectively symmetric superperiod is shown in Figure 1. Two 28.5° bends and a single central 12° reversed bend yield an overall bending angle of 45°. Quadrupoles are symmetrically excited in two families to provide horizontal focusing; dipoles, similarly, have gradients to give vertical focusing. Quad strengths, dipole gradients, and drift lengths are selected to provide betatron stability with a horizontal tune of ¾, a vertical tune of ¼, achromaticity, and isochronicity.

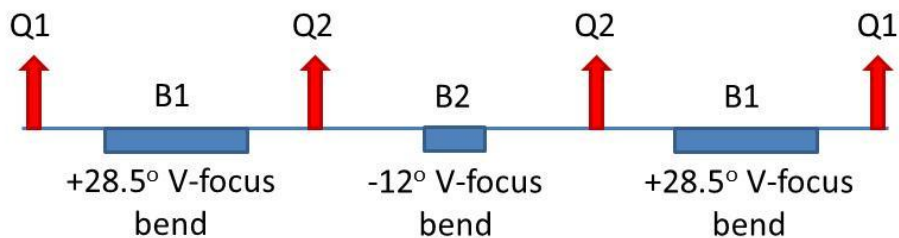

Figure 1: Achromatic, isochronous TBA superperiod structure. Two quad and two dipole (gradient) families provide horizontal and vertical tunes of ¾ and ¼, respectively, with overall zero dispersion and momentum compaction.

TBAs typically use dispersion modulation for compaction control, and tend as a consequence to be strongly focusing and highly chromatic. In this case, the isochronicity constraint is readily met through use of the reversed direction of the small-angle central bend; strong focusing is unnecessary and severe chromatic effects are avoided. Each superperiod is individually chromaticity-corrected and driven to $T_{566}$=0 through use of sextupole terms embedded in the quads and dipoles. When four superperiods are concatenated to form an 180° arc (Figure 2), the overall structure is a unit matrix transform – and achromatic and isochronous – through second order. Transport parameters (Figure 3) are insensitive to energy (Figure 4). Table 1 provides a list of key lattice parameters.

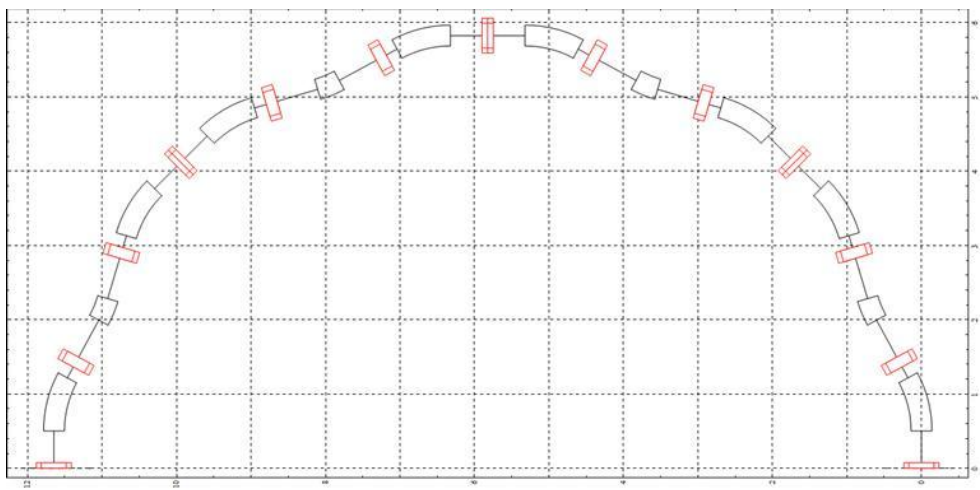

Figure 2: Full 180° arc of four TBA superperiods.

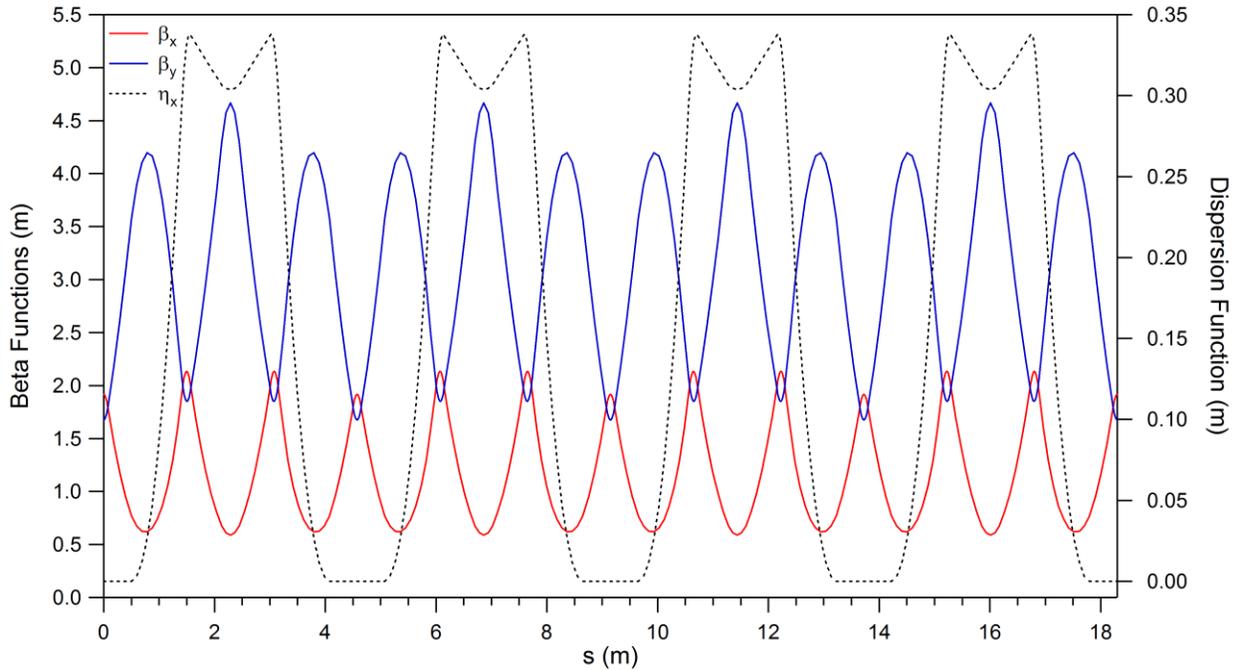

Figure 3: Twiss parameters and dispersion for full arc.

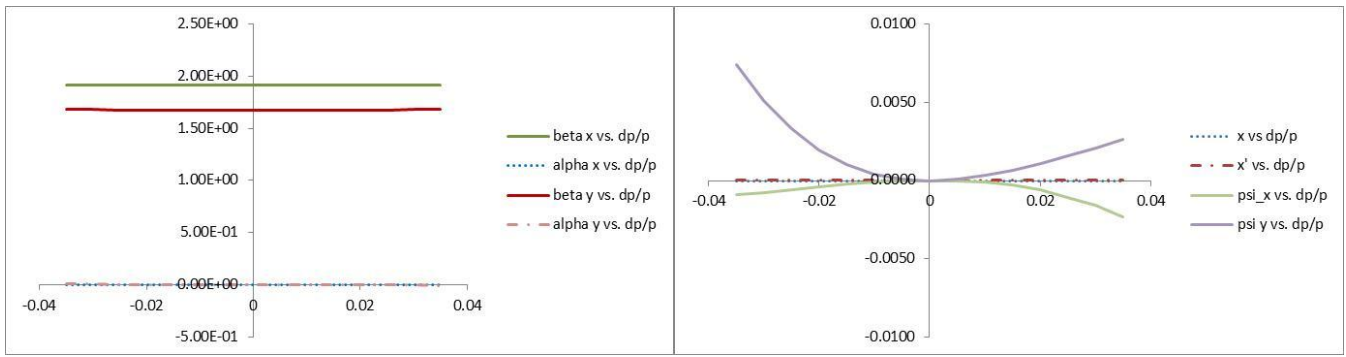

Figure 4: Result of momentum scan of full arc. Left: Twiss parameters as function of relative momentum offset; right: bend plane orbit and transverse phase advances as function of relative momentum offset.

Table 1: Key parameters of TBA-based isochronous arc

| | | | |
|---|---|---|---|
| Energy range | 200-600 MeV | Superperiod dispersion $\eta_x$, $\eta'_x$ | 0 m, 0 rad |
| Superperiod length | 4.575 m | Superperiod compactions $R_{56}$, $T_{566}$ | 0 m, 0 m |
| Bend radius | 1.5 m | Arc structure | 4 superperiods |
| Bend angles | 28.5°, −12°, 28.5° | Arc length | 18.3 m |
| Superperiod tune $\nu_x$, $\nu_y$ | ¾, ¼ | Average arc radius | 5.825 m |
| Matched $\beta_x$, $\beta_y$ | 1.917, 1.678 m | Arc geometric angle | 180° |
| Superperiod chromaticity $\zeta_x$, $\zeta_y$ | 0, 0 | Arc tune $\nu_x$, $\nu_y$ | 3, 1 |

The performance of this arc in the presence of CSR was simulated using `elegant` [5] at an energy set point of 200 MeV. The evolution of the momentum compaction (transport matrix element $R_{56}$) will be relevant to subsequent discussion, and is thus shown in Figure 5. A bunch charge of 150 pC was assumed to have transverse and longitudinal emittances of 0.5 mm-mrad and 35 keV-psec, respectively. The longitudinal phase space was chirped in order to simulate conditions typical during transport of a beam in a multipass SRF ERL, such as JLAMP [6], with rms length $\sigma_z$ = 3.0 psec and rms momentum spread $\sigma_{\delta p}$ = 11.67 keV.

Figure 6 presents the evolution of the transverse emittances. No degradation is observable, despite centroid energy loss to CSR (Figure 7). We note that losses to incoherent synchrotron radiation (ISR) are, at 200 MeV, quite small. Of equal interest is the observed phase space behavior: both transverse (Figures 8a, 8b) and longitudinal (Figure 8c) phase spaces are – at the end of the arc – completely undistorted images of the input distributions. Of additional interest is the absence of any obvious microbunching of the longitudinal phase space.

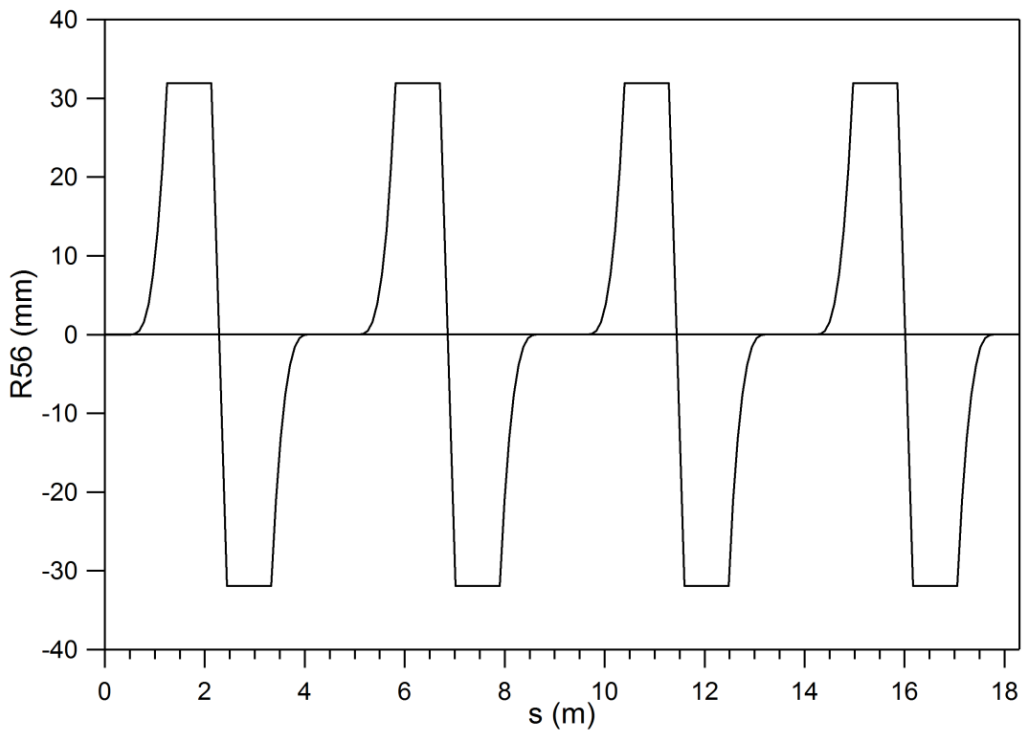

Figure 5: Evolution of $R_{56}$ through TBA-based isochronous achromat.

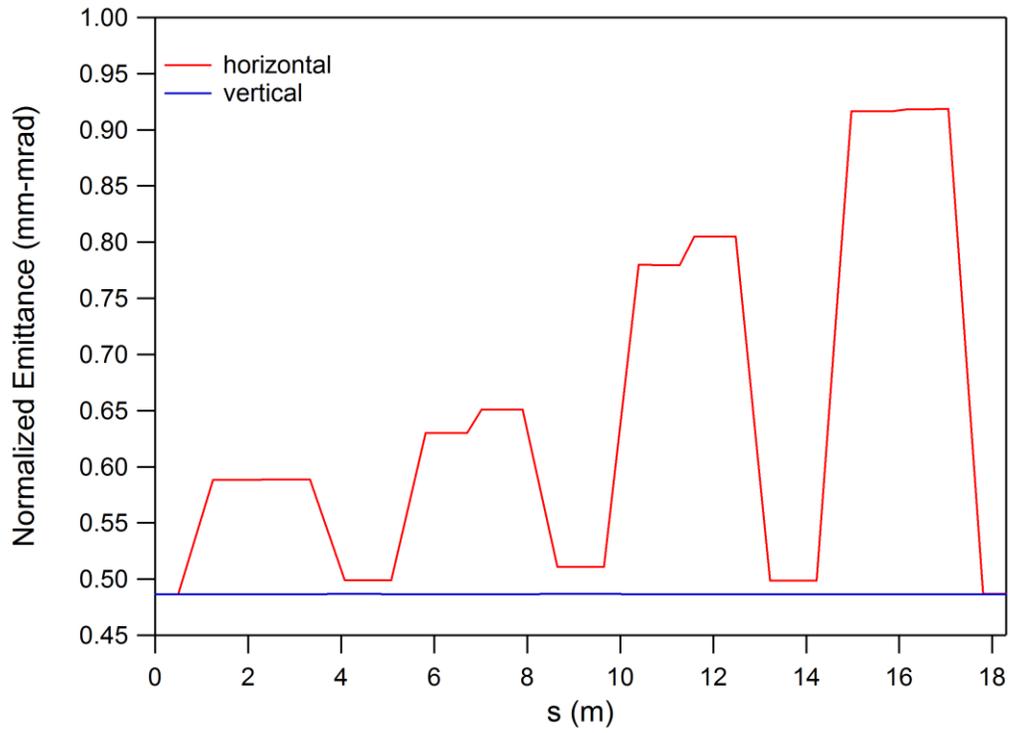

Figure 6: Evolution of transverse normalized emittance through arc.

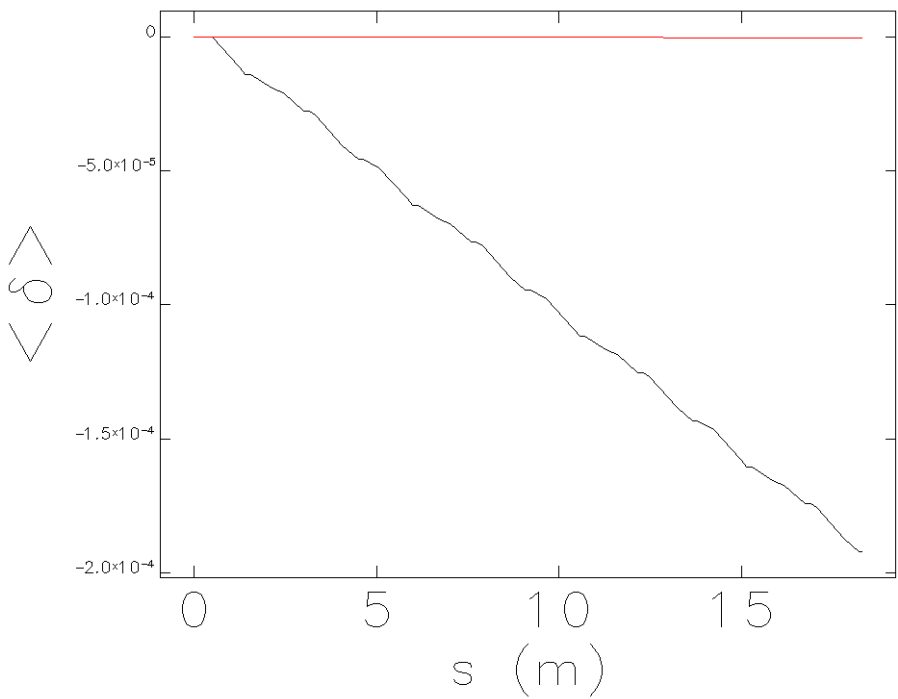

Figure 7: Energy loss due to ISR (red) and CSR+ISR (black) during transport through arc.

The complete absence of any observable effect from CSR (other than energy loss) suggests that optics balance across the achromat provides the expected emittance compensation. The absence of microbunching is similarly intriguing, and will be the subject of additional discussion below.

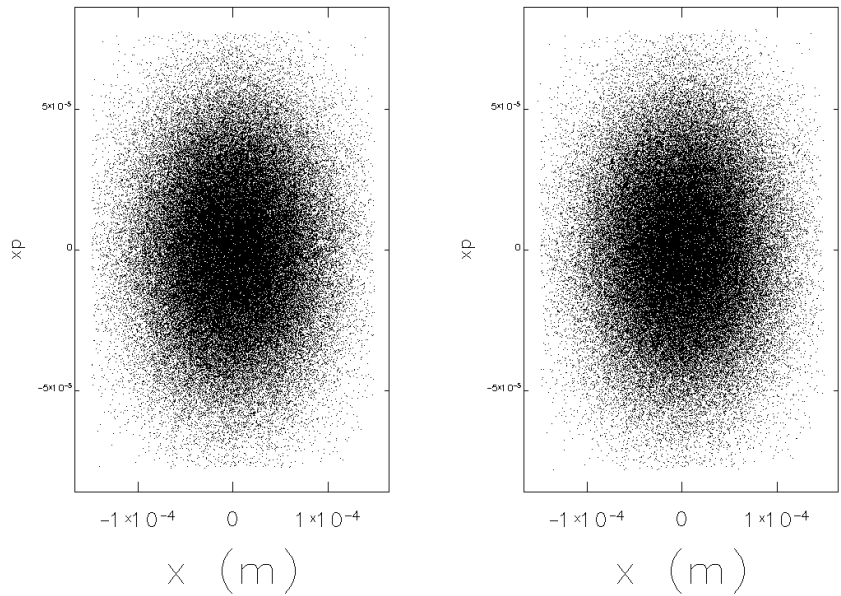

Figure 8a: Bend plane phase space at entrance (left) and exit (right) of arc.

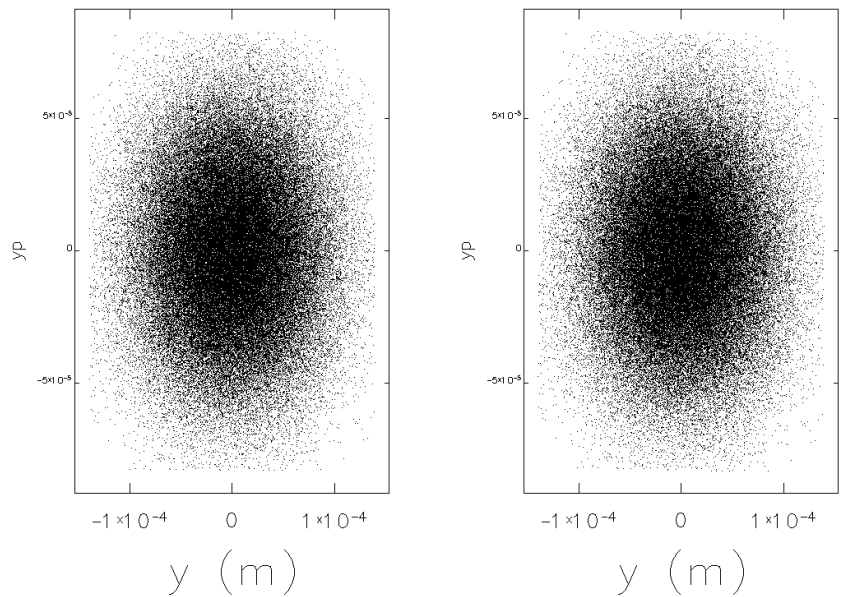

Figure 8b: Non-bend plane phase space at entrance (left) and exit (right) of arc.

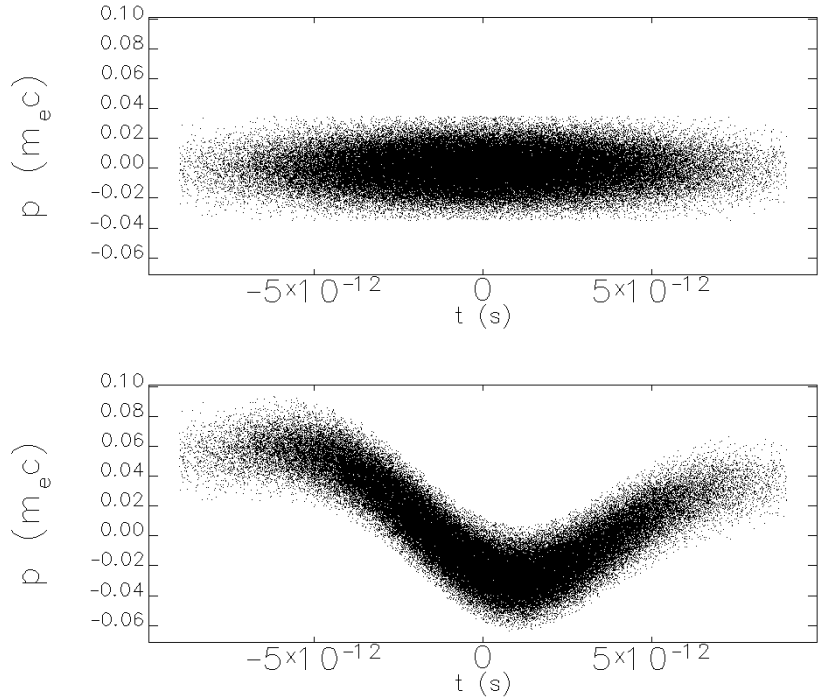

Figure 8c: Longitudinal phase space at entrance (left) and exit (right) of arc.

## Example 2: High Energy Recirculation Arc

Nearly all requirements for CSR suppression in a high-energy recirculator are met by the original design concept for the CEBAF arcs [7]. Control can be enhanced by adding provisions for small bend-plane beam envelope in the dipoles and provision for control of terms such as $T_{566}$ (though this is in principle possible with only minor modification of the "stock" CEBAF transport system [8]).

Given a desire for stringent emittance control and flexibility in setting the longitudinal match, we have therefore generated a slightly modified version of the CEBAF arc transport line. Figure 9 presents a conceptual representation of one of the four superperiods of a CEBAF arc. It is simply a pair of back-to-back 90° FODO dispersion suppressors; it is nominally linearly achromatic, imaging in both transverse planes (M ~ I), but nonisochronous. By increasing the strength of the highlighted quadrupole pair (which are separated by 180° in betatron phase), the dispersion is driven down in the inner pair of dipoles (reducing the momentum compaction), and the horizontal and vertical tunes split (horizontal upward, vertical downward). Fitting on all quad families then allows a precise trim of matched beam envelope, achromaticity, momentum compaction, and tune. A choice of quarter integer tunes (5/4 horizontal, 3/4 vertical) then leads to the desired second order achromatic configuration when four superperiods are used to generate a complete arc.

This configuration is disadvantageous in that the bend-plane beam envelopes are not forced to small values in the dipoles. This increases betatron response to radiation events – aggravating both ISR and CSR, and decreases beam divergence in the dipoles, magnifying the relative impact of a radiative shift in dispersive angle. In addition, control of $T_{566}$ and other nonlinear compaction terms is not entirely transparent.

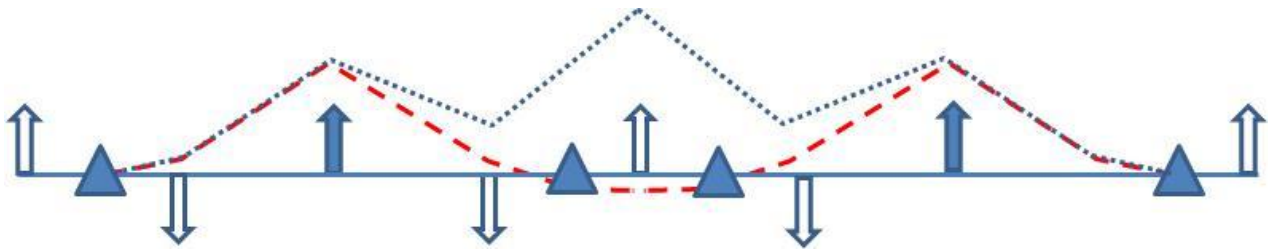

Figure 9: CEBAF superperiod, with notional dispersion pattern for initial tuning as paired 90°
FODO dispersion suppressors (blue dotted) and – after increase in strength of highlighted quads
– for quarter-integer isochronous achromat (red dashed).

In order to reduce in-bend bend-plane envelopes and provide an "orthogonal knob" for $T_{566}$, we
have in this study used a slightly different focusing structure, based on "an achromat derived from an an
achromat". We begin by generating a single 90° "theoretical minimum emittance" (TME) focusing
structure [9] as shown in Figure 10. When four such cells are put together, an achromatic (to second
order) but nonisochronous superperiod results (Figure 11). By increasing the strength of the highlighted
quadrupoles (which have 180° betatron phase separation), the dispersion can be driven down in the
inner dipoles, the tunes split, and a linearly achromatic, isochronous superperiod obtained. As with
CEBAF, optimization using all quad families then allows choice of tune, matched envelopes, enforced
achromaticity, and selection of momentum compaction.

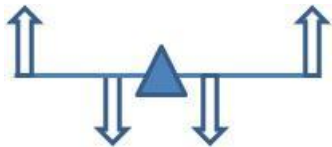

Figure 10: Single period building block of modified transport line.

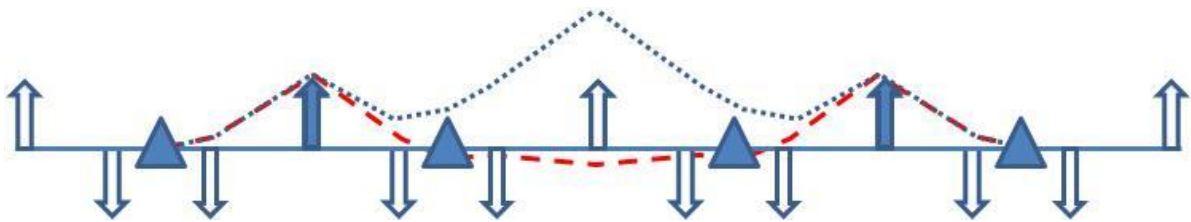

Figure 11: Modified superperiod, with notional dispersion pattern for initial tuning as four 90°
TME cells in second order achromat (blue dotted) and – after increase in strength of highlighted
quads – for quarter-integer isochronous linear achromat (red dashed).

After simulation trials of solutions using different superperiod tunes, we find that sixth-integer tunes
(7/6 horizontal, 5/6 vertical) provide good chromatic behavior and admits a particularly simple means of
control of $T_{566}$ (and in principle $W_{5666}$), as discussed below. Six superperiods then form a second-order
achromatic arc; the higher periodicity also reduces the individual bend angle below that used in CEBAF,
providing further mitigation of radiation effects. Figure 12 shows beam envelope functions for a full superperiod. As noted horizontal beam envelopes are small in the bends. As there are a total of 24 bends per arc, each dipole provides 7.5° deflection, yielding small η' in the bends and – together with the relatively large beam divergence at the same location – helping reduce the impact of both ISR and CSR. Moreover, the small beam envelopes and very small dispersion in all dipoles (a result of achromaticity and the dispersion modulation imposed to make each superperiod isochronous) reduces response to CSR and ISR; note that the excitation function $\langle \mathcal{H} \rangle = 0.0454$ m so that ISR is well controlled. Table 2 provides a list of relevant lattice parameters.

Table 2: Key parameters of TME-based isochronous arc

| Energy | 1.3 GeV | Superperiod dispersion $\eta_x$, $\eta'_x$ | 0 m, 0 rad |
| --- | --- | --- | --- |
| Superperiod length | 40 m | Superperiod compactions $R_{56}$, $T_{566}$ | 0 m, 0.878 m |
| Bend radius | 3.614 m | Arc structure | 6 superperiods |
| Bend angle | 7.5° | Arc length | 240 m |
| Superperiod tune $\nu_x$, $\nu_y$ | 7/6, 5/6 | Average arc radius | 76.3 m |
| Matched $\beta_x$, $\beta_y$ | 65, 2.5 m | Arc geometric angle | 180° |
| Superperiod chromaticity $\zeta_x$, $\zeta_y$ | 0, 0 | Arc tune $\nu_x$, $\nu_y$ | 7, 5 |

Most importantly, the betatron phasing associated with the second order achromatic architecture also introduces emittance compensation in the manner discussed above: each dipole has a partner a half-betatron-wavelength away, at which the bunch length and all beam envelope functions are the same, so that emittance-degrading effects cancel. This is particularly strongly enforced by use of a periodically isochronous structure – which insures that the bunch length is the same at phase-homologous CSR emission sites.

The small dispersion and dispersive slope result not only in a small (zero) momentum compaction in each superperiod, the modulation of the momentum compaction through the system is extremely small, potentially providing some limitation on microbunching gain. Figure 13 shows the evolution of $R_{56}$ through a superperiod; max/min values are at the millimeter level.

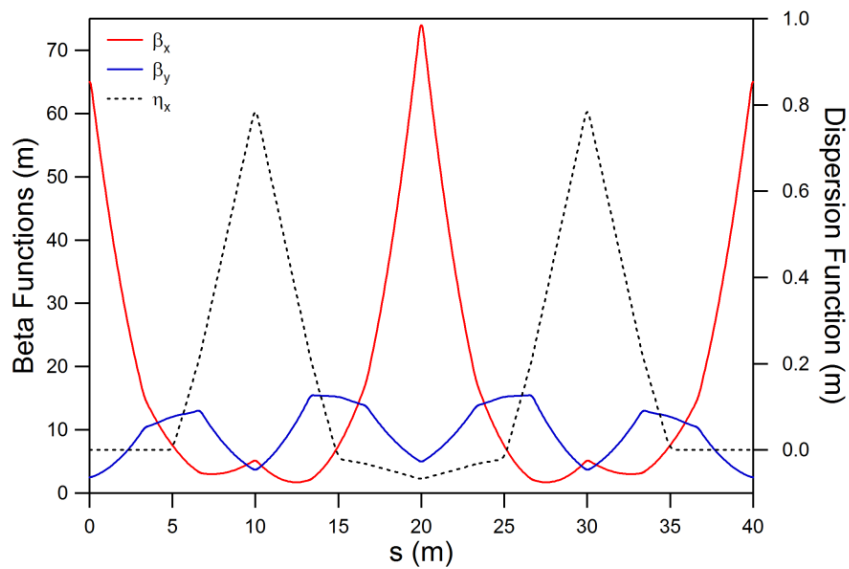

Figure 12: Lattice functions for single superperiod relative to beamline elements.

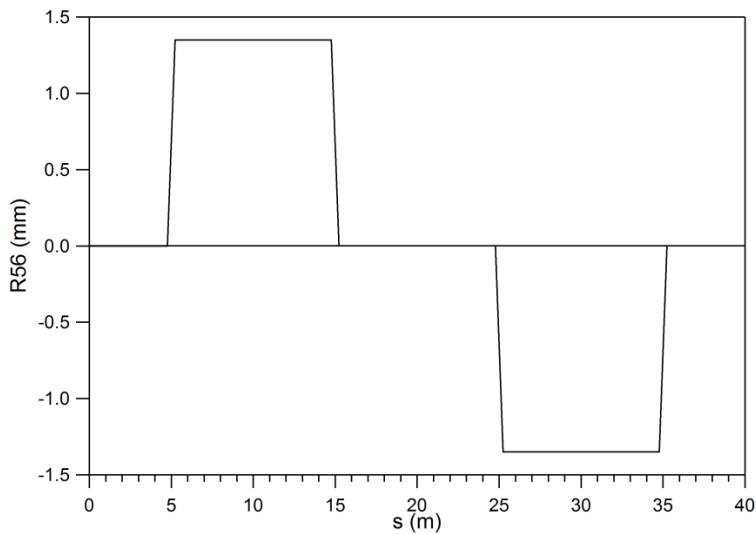

Figure 13: Evolution of $R_{56}$ through single superperiod.

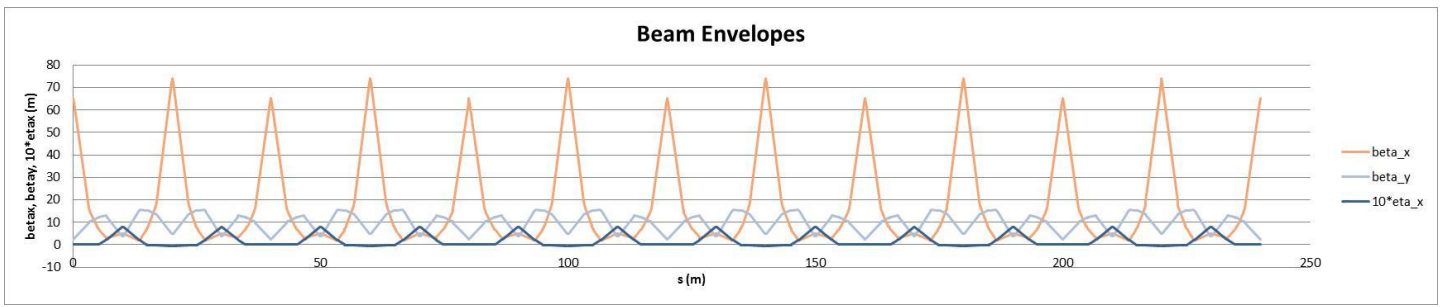

Figure 14 shows beam envelope functions for an entire arc (six superperiods). Periodicity is evident; this simplifies implementation of chromatic correction, discussion of which follows.

Chromatic correction is – as in conventional systems – intended to control emittance degradation due to aberrations, assist in instability management, and alleviate sensitivity to machine/beam energy drifts. In this case, it is also intended to support control curvature of the beam longitudinal phase space – often a desirable feature of a recirculator longitudinal match.

We have adopted a simple solution in which one family of sextupoles is adjusted to generate a nonlinear dispersion bump – creating a desired value for $T_{566}$ (in this case, ~5 m); two other families are then set to render the entire arc a second order achromat. Figure 15 presents nomenclature for the quadrupole/sextupole families that are used. Various combinations of families (with and without reflective symmetry) were tested until a solution with very good performance was obtained. Q7 provides the nonlinear dispersion bump and is manually adjusted to provide a trial solution; sextupole components in Q3/Q3X and Q4/Q4X are then set (by numerical optimization) to zero the horizontal and vertical chromaticities of the superperiod. This process was iterated until the desired $T_{566}$ value was obtained. All elements in the resulting second order transformation matrix are thus zero, save for the deliberate offset in $T_{566}$.

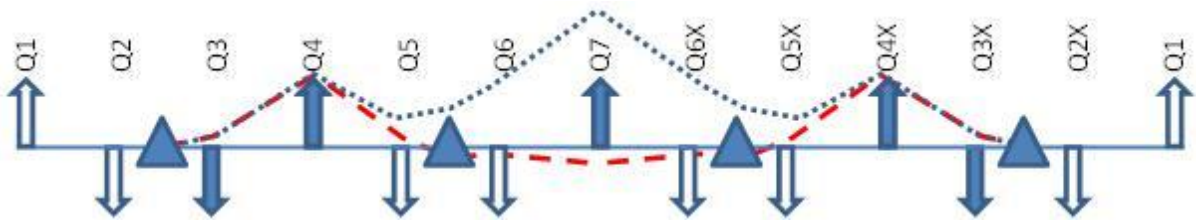

Figure 15: Nomenclature for chromatic correction. Sextupole components in highlighted elements are used for correction; Q7 provides $T_{566}$ control, Q3/Q3X and Q4/Q4X correct chromaticity.

Performance is quite good. Figure 16 gives the results of momentum scan and indicate the system has in excess of 2% full momentum acceptance. Figure 17 presents a geometric aberration analysis showing that the system is benign out to amplitudes of 100 times the nominal emittance. Nonlinear detuning (Figure 18) is modest.

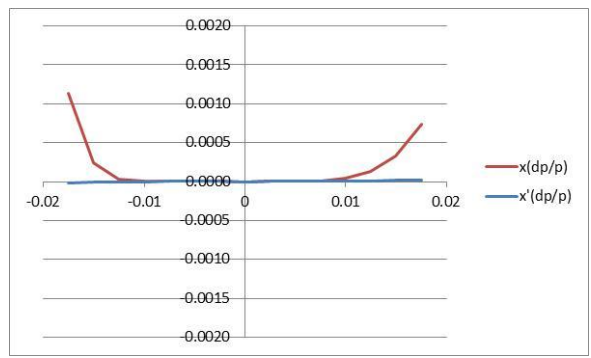

Figure 16a: Momentum scan of bend plane orbit dependence on momentum offset.

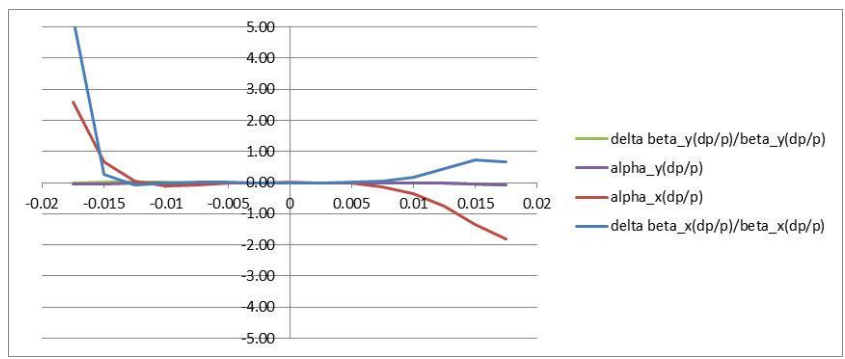

Figure 16b: Momentum scan – variation of Twiss parameters with momentum offset.

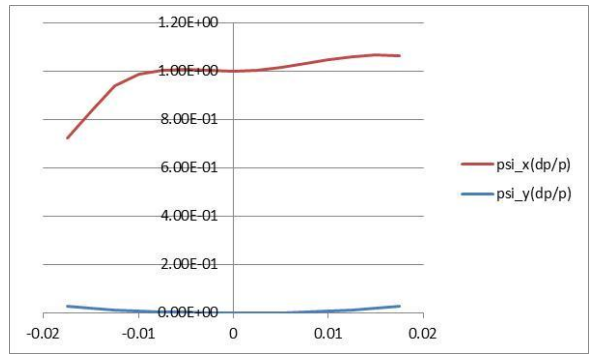

Figure 16c: Momentum scan – phase advance as a function of momentum offset.

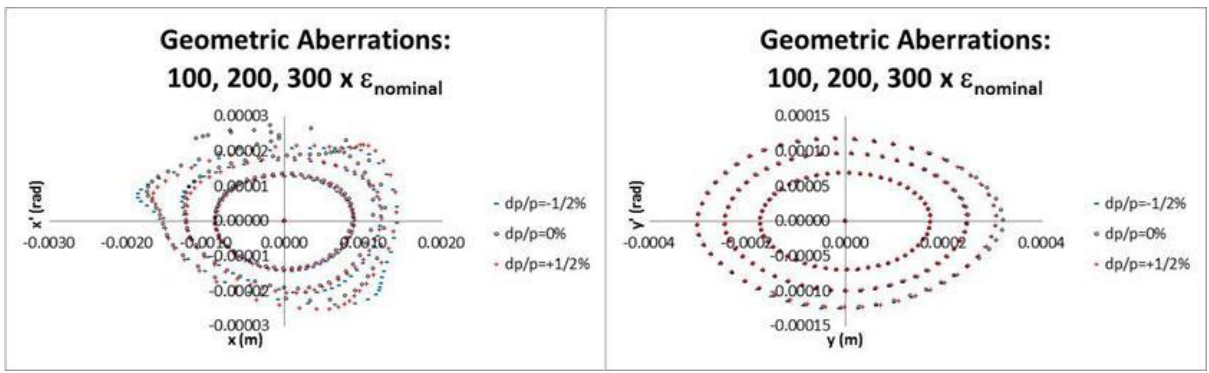

Figure 17: Geometric aberration analysis; phase space is regular out to beyond 100 times nominal emittance, even at ±½% momentum offsets.

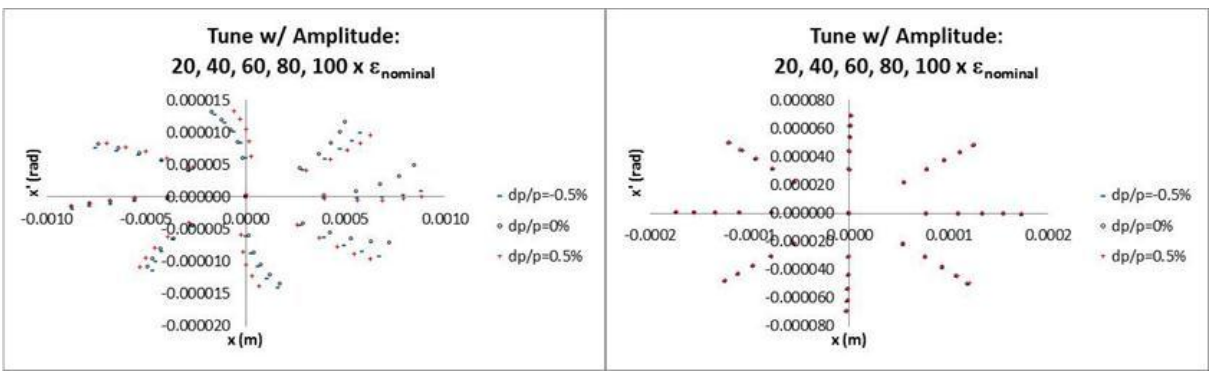

Figure 18: Nonlinear detuning analysis presenting only modest change in phase advance with amplitude.

Overall performance with respect to aberrations is thus quite good, especially for such a simple system. Additional control can – at least in principle – be imposed through use of higher order correction elements such as octupoles or decapoles, but even the simple solutions presented above provide beam quality preservation adequate to proceed to an analysis of the impact of radiation effects.

*CSR simulation* – As in Example 1, the bare lattice is benign and has little impact on beam quality. Simulation of transport with CSR was performed with `elegant` at 1.3 GeV. We find that the emittance compensation provided by optics balance across the achromatic is excellent. Figure 19 shows the evolution of an initial transverse normalized emittance of 0.25 mm-mrad emittance through the line for different bunch charges, assuming an initially upright bunch of length of 3 psec and momentum spread 11.67 keV. The emittance compensation is nearly perfect; final transverse phase spaces appear nearly identical to initial (Figure 20). Emittance is well-preserved over a broad range of bunch charges and initial bunch lengths (Figure 21), despite the loss of energy via CSR (Figure 22).

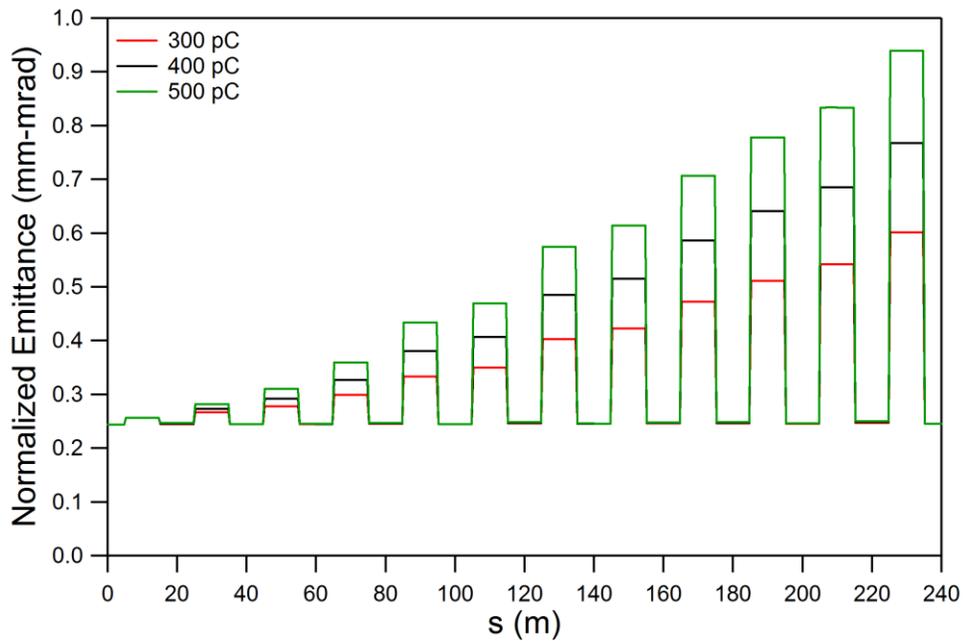

Figure 19: Horizontal normalized emittance evolution through the arc for bunch charges of 300, 400 and 500 pC.

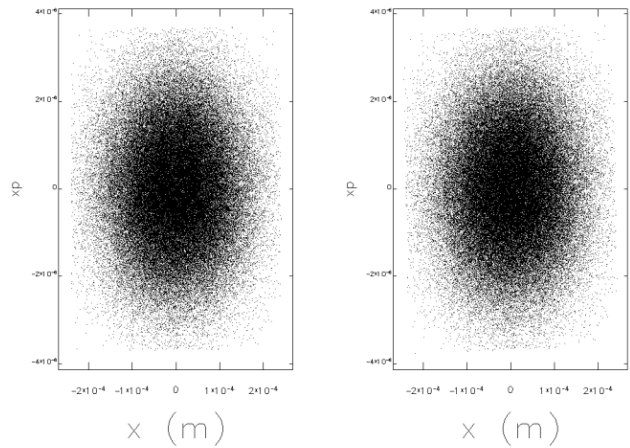

Figure 20a: Horizontal phase space before (left) and after (right) transport at 300pC.

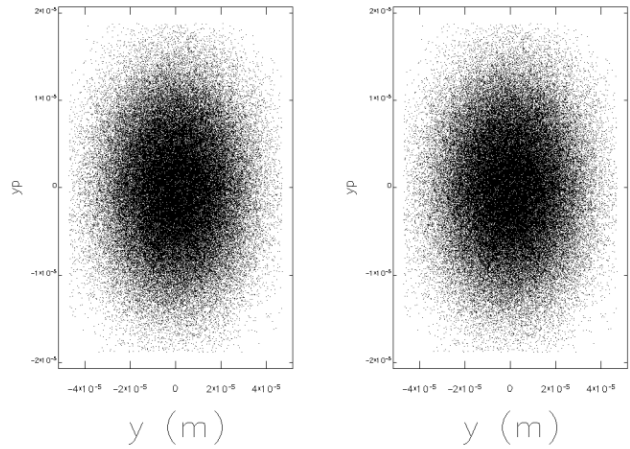

Figure 20b: Vertical phase space before (left) and after (right) transport at 300 pC.

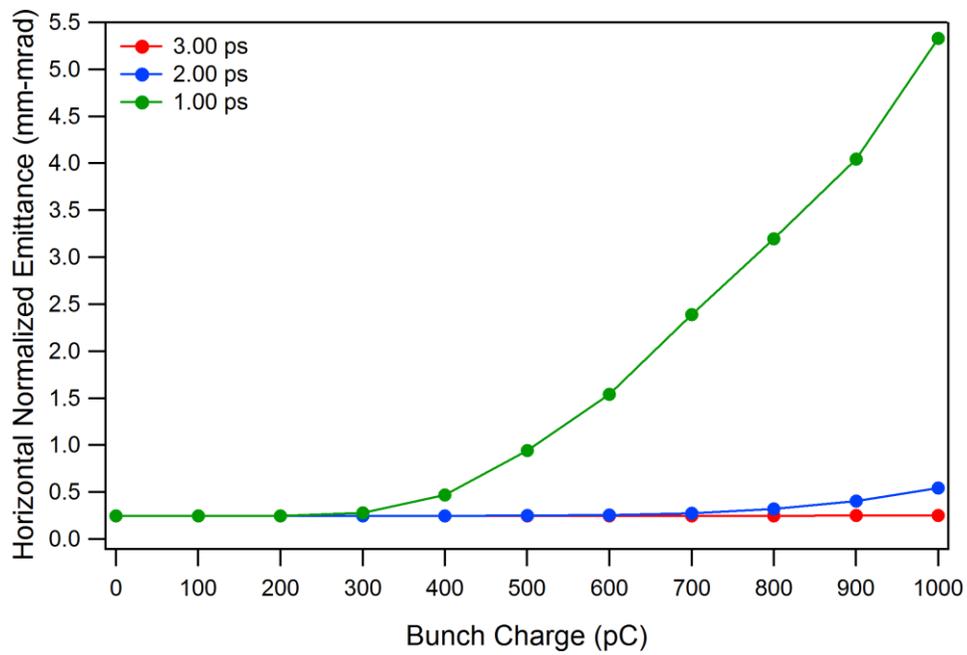

Figure 21: Final emittance vs. bunch charge at various RMS bunch lengths; up to 700 pC virtually no increase is observed for bunch lengths of order 2 psec or longer.

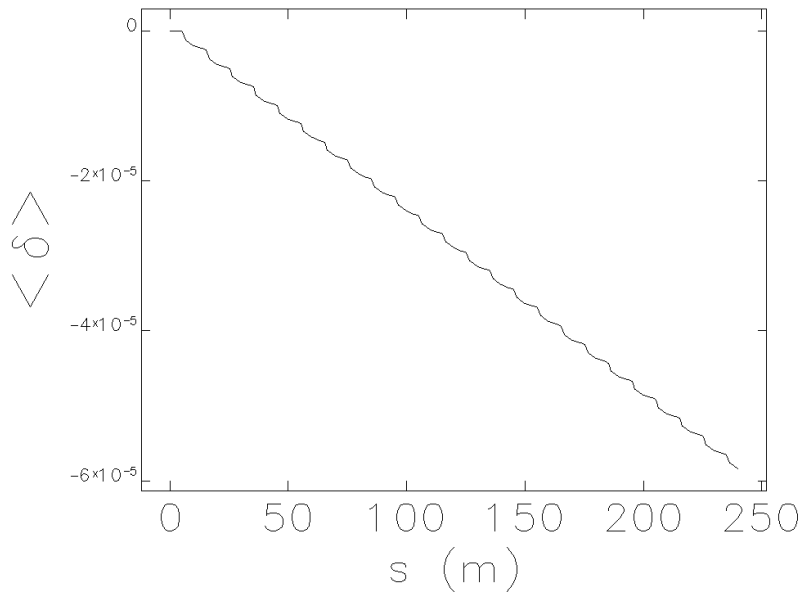

Figure 22: Energy loss to CSR and ISR across arc at 300 pC.

Of considerable interest – as in Example 1 – is the preservation of longitudinal beam quality, and in particular the behavior of microbunching effects. Figure 23 shows the longitudinal phase space after 1.3 GeV transport – at various charges – of a bunch with 0.25 mm-mrad transverse emittance, 3 psec rms bunch length and an rms relative momentum spread of $8.974 \times 10^{-6}$. Though wake distortion is evident, there is no visible microbunching until the bunch charge approaches 1 nC.

We remark that although the wake distortion is significant, absent microbunching it is analogous to the impact of RF curvature and may be compensated using similar methods. In particular, it is possible to correct such distortions through use of an appropriate sequence of chirps and magnetic nonlinear compactions during acceleration and transport through a multipass recirculated linac.

**Microbunching Instability Suppression**

Simulation results of both example arcs not only exhibit no transverse emittance dilution from CSR; neither arc, in addition, manifests evidence of the μBI unless the transported bunch is quite short or the charge quite high (several 100s of pC). Given discussions in the literature [10], we suspect that the apparent low microbunching gain is due to the small amplitude of $R_{56}$ oscillations along the periodically isochronous arc. To test this conjecture, we have therefore performed a simulation experiment in which a high energy arc – quite similar to that about – was configured so to exacerbate microbunching. The microbunching gain for both arcs was then evaluated and compared; discussion of these exercises is the subject of the remainder of this note.

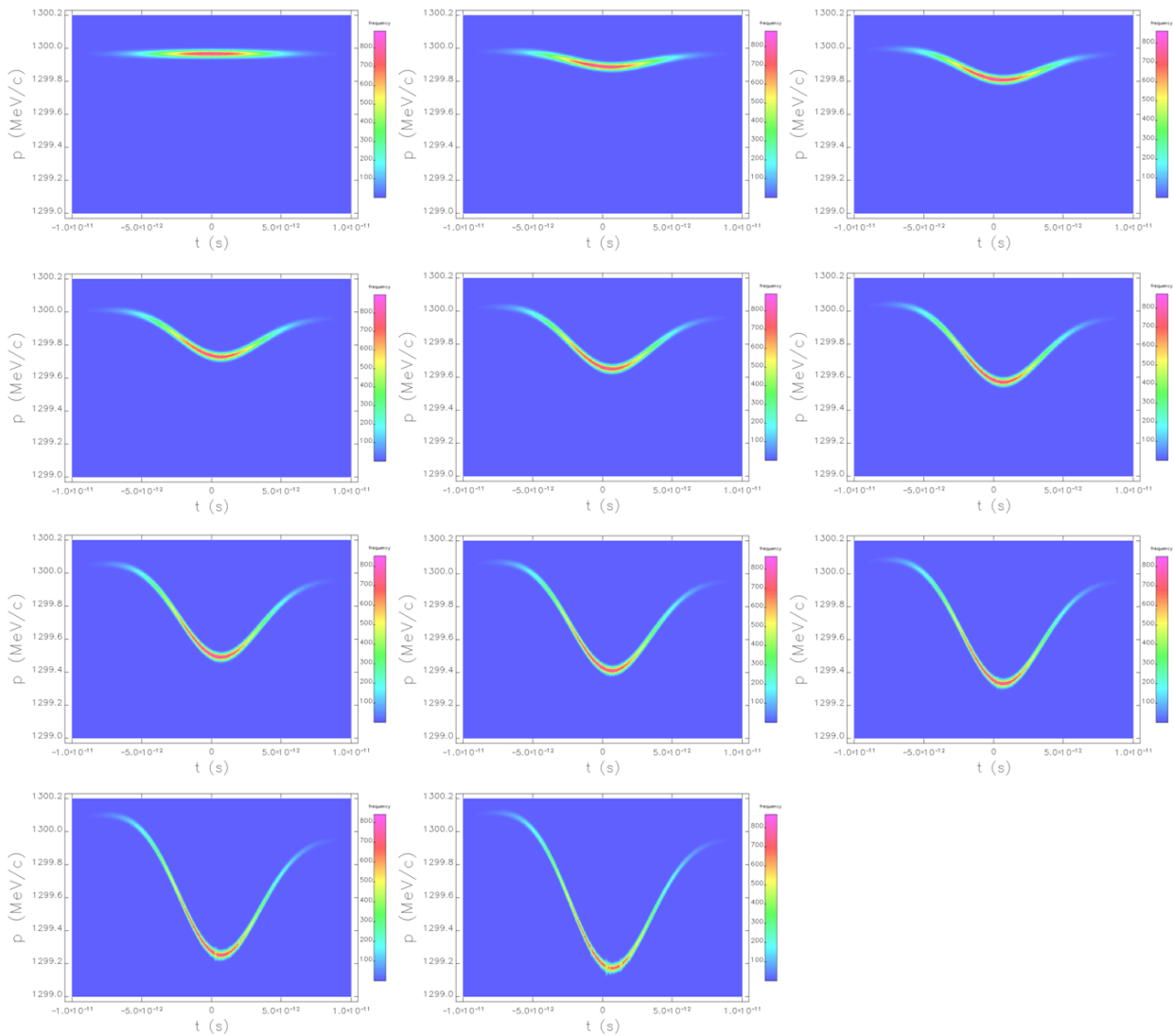

Figure 23: Longitudinal phase space at the exit of the arc for bunch charges from 0 pC to 1000 pC (from left to right and top to bottom) in steps of 100 pC. First signatures of microbunching structure occur at 900 pC.

## Example 3: Alternate High-Energy Recirculation Arc: 2nd Order Achromat With Large Dispersion/Large $R_{56}$ Oscillations

To explore the effectiveness of the periodically isochronous/2nd order achromat-based CSR/mBI suppression scheme, we have generated a third example design. It is based on Example 2, and is also a second order achromat, but rather than individually isochronous superperiods, it is rendered globally isochronous by dispersion modulation across the entire arc. This allows an assessment of the impact of large compaction oscillations, large dispersion, and the absence of multiply periodic isochronicity.

Starting with the same basic TME cell and four-cell superperiod structure employed in Example 2, a process similar to that used in early CEBAF design studies [11] is used. We tune each TME cell to fractional tunes of 5/24 horizontally and vertically (instead of splitting tunes). Four TME cells then form a superperiod with tunes of 5/6 in both planes; quads in each superperiod are fit to hold this tune while forcing $R_{56}$ to zero. Individual superperiods are, however, not achromatic, by virtue of the weaker horizontal focusing/lower horizontal tune. Though the constraint that $R_{56}=0$ locally eliminates the linear dependence of path length on energy, superperiods are thus not strictly isochronous. When six such superperiods are combined to form a complete 180° arc, the resulting second order achromat then displays an oscillatory dispersion pattern that in turn drives the arc to be achromatic from end to end and the suppresses the overall momentum compaction. A final fit of the arc as a whole renders it six-fold periodic, sets the tunes to 5 wavelengths in both planes, holds achromaticity, and forces the line to be isochronous from end to end.

Individual superperiods remain at 5/6 integer tunes, but have nonzero matched dispersion and small but nonzero (positive) $R_{56}$. Matched Twiss parameters are modest; the large dispersion oscillation reaches a peak of ~4 m amplitude. Beam envelopes, dispersion, and the strongly oscillatory evolution of $R_{56}$ are shown in Figure 24. Table 3 provides a list of key lattice parameters.

Table 3: Key parameters of TME-based high/oscillatory-dispersion arc

| Energy | 1.3 GeV | Superperiod dispersion $\eta_x$, $\eta'_x$ | -1.601 m, 0 rad |
|---|---|---|---|
| Superperiod length | 40 m | Superperiod $R_{56}$, $T_{566}$ | 0.062 m, 0.240 m |
| Bend radius | 3.614 m | Arc structure | 6 superperiods |
| Bend angle | 7.5° | Arc length | 240 m |
| Superperiod tune $\nu_x$, $\nu_y$ | 5/6, 5/6 | Average arc radius | 76.3 m |
| Matched $\beta_x$, $\beta_y$ | 35.87, 3.00 m | Arc geometric angle | 180° |
| Superperiod chromaticity $\zeta_x$, $\zeta_y$ | 0, 0 | Arc tune $\nu_x$, $\nu_y$ | 5, 5 |

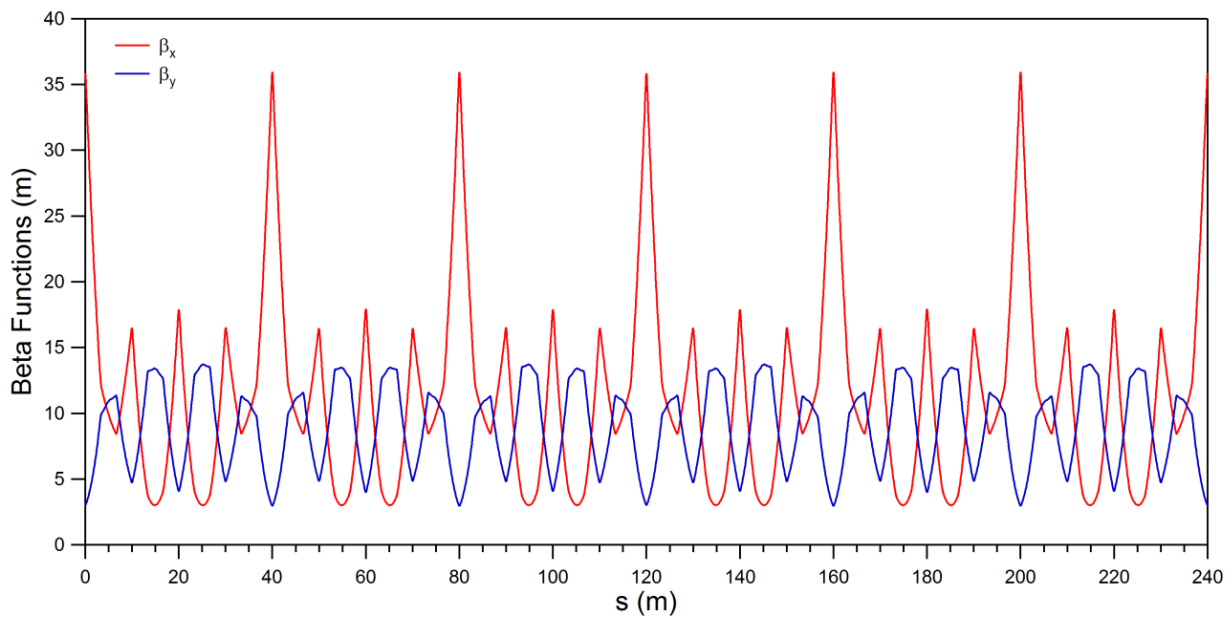

Figure 24a: Beam envelopes/dispersion for high dispersion tuning.

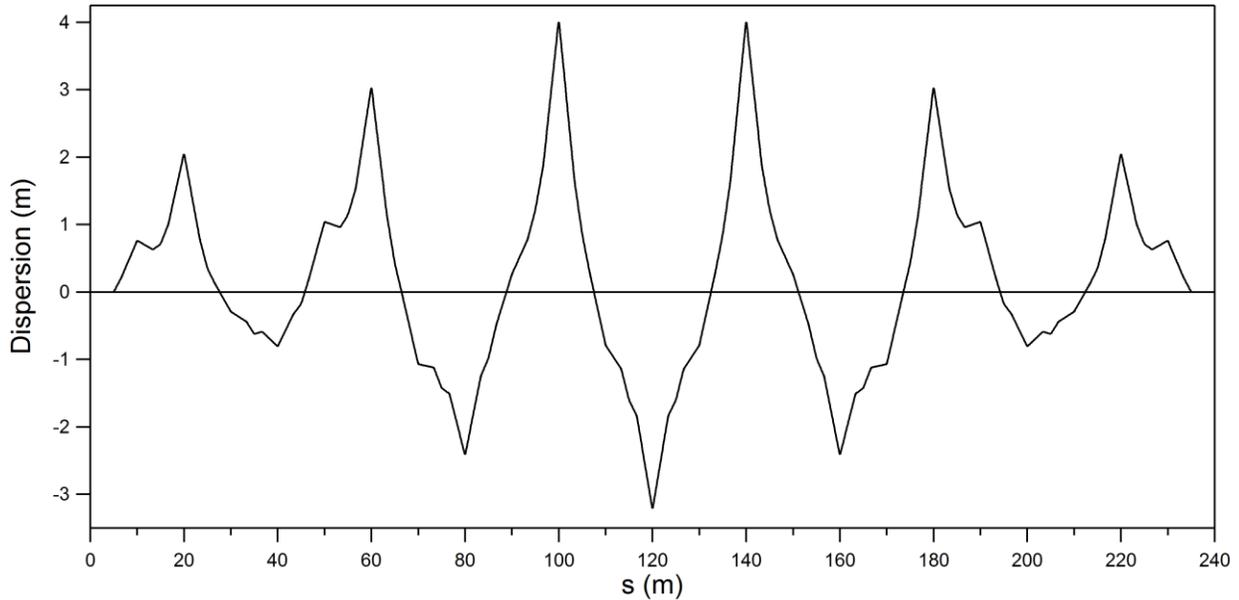

Figure 24b: Dispersion through full arc for high-dispersion tuning

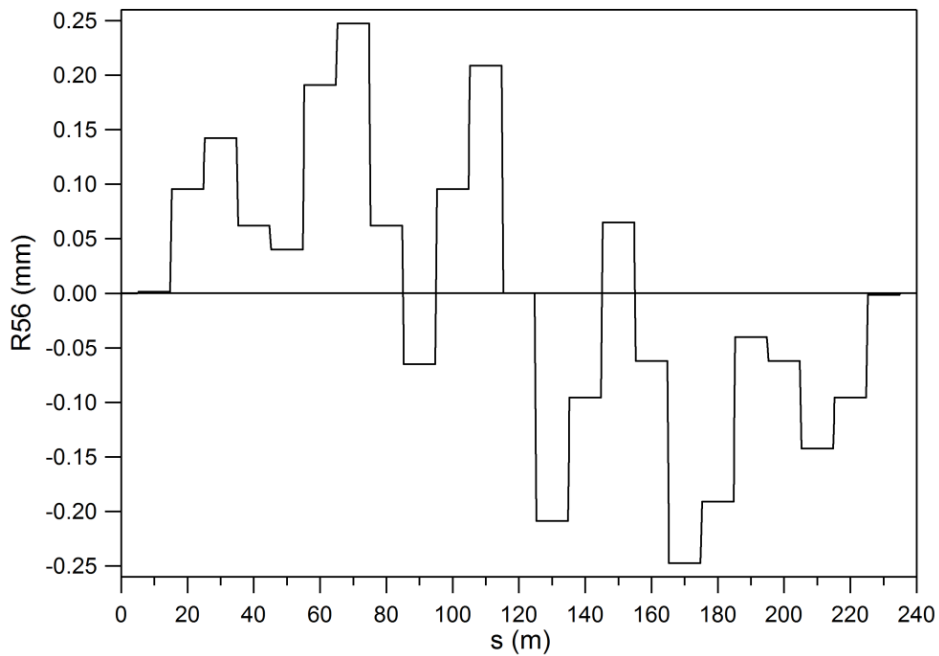

Figure 24c: Aperiodic, large-amplitude evolution of $R_{56}$ through high-dispersion arc of Example 3.

Chromatic correction – as in Example 2 – is accomplished using typical methods. In the notation of Figure 15, sextupole terms in two quad families (Q5/Q5X and Q7) are used to correct the chromaticities of each superperiod, according to the usual procedure for a second order achromat. A second order dispersion bump is then imposed by a small trim on the strength of the sextupole term in the Q7 of the second and fifth superperiod, which are of course separated in phase by 2½ betatron wavelengths. A small sextupole trim in Q1 is applied to manage nonlinear dispersive and focusing effects, and the process iterated until results are satisfactory. Figure 25 presents the results of momentum scans; behavior is – as in Example 2 (Figure 16) – quite good. Figure 26 presents a geometric aberration analysis showing that the system is benign out to amplitudes of 300 times the nominal emittance. Nonlinear detuning (Figure 27) is – as in Example 2 – modest.

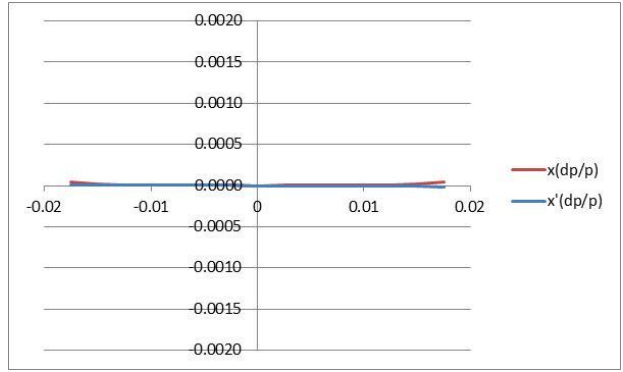

Figure 25a: Momentum scan – bend plane orbit dependence on momentum offset in high dispersion achromat.

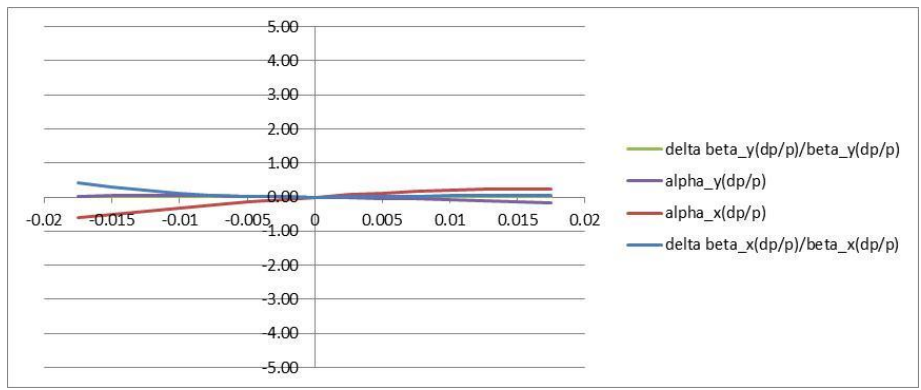

Figure 25b: Momentum scan – variation of Twiss envelope parameters with momentum offset in high dispersion achromat.

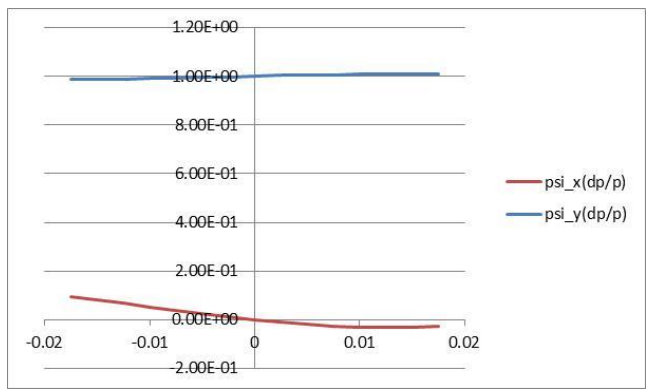

Figure 25c: Momentum scan – phase advance as a function of momentum offset in high dispersion achromat.

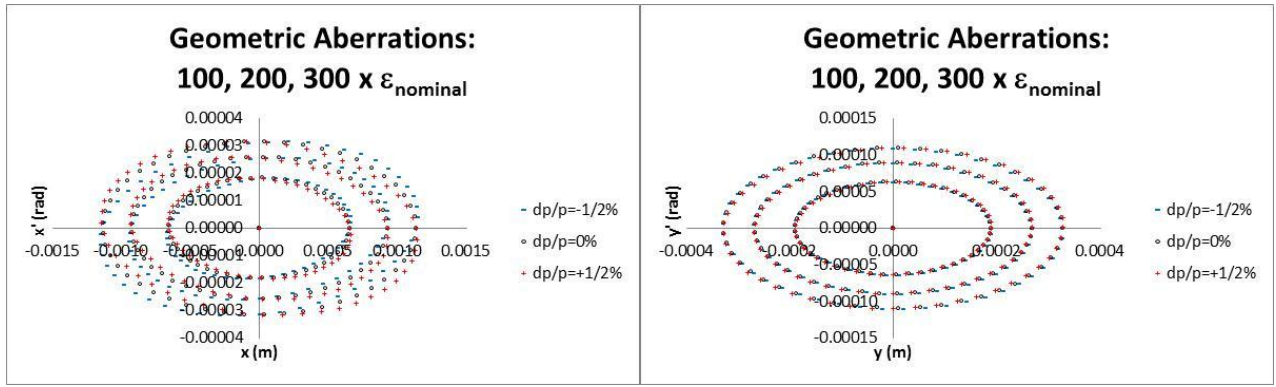

Figure 26: Geometric aberration analysis for high dispersion achromat; phase space is regular out to beyond 300 times nominal emittance, even at ±½% momentum offsets.

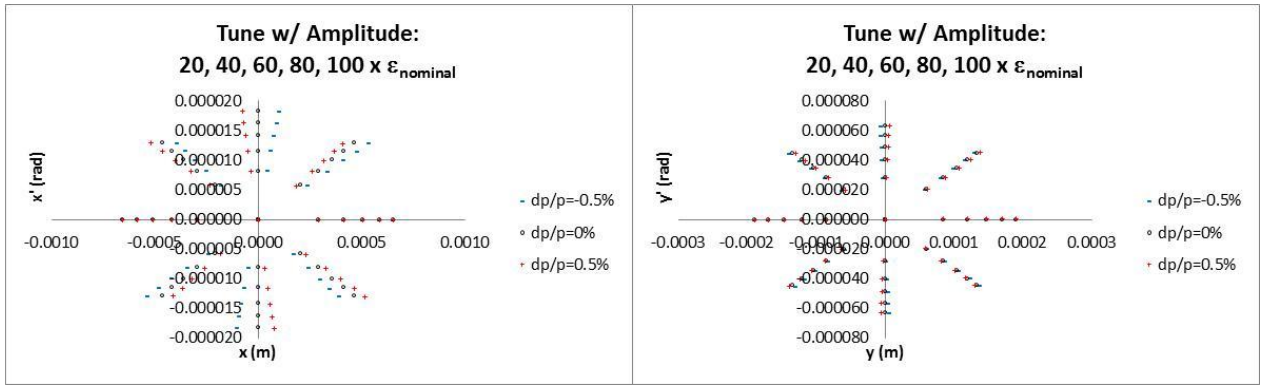

Figure 27: Nonlinear detuning analysis presenting only modest change in phase advance with amplitude.

_CSR simulation_ – As in Example 2, the bare transport lattice has essentially no impact on beam quality; simulation of idealized many-particle distributions show no beam quality degradation in the absence of CSR. Simulation with CSR using elegant reveals, however, an enormous difference from Example 2. In this case, emittance growth is not well-compensated (Figure 28) and microbunching is quite evident at charges well below those needed to excite the instability in Example 2 (Figure 29).

The contrast with Example 2 results is striking, and suggests again that the emittance compensation method is an effective means of CSR control, and it moreover provides this control in a manner that tends to suppress the microbunching instability. This breakdown in emittance compensation and exacerbation of microbunching effects appears to be related – as suggested above – to the magnitude of the R56 oscillations, as well as to the absence of symmetry/dispersive periodicity in this lattice. We have therefore evaluated the microbunching gain in for each of the Example 2 and 3 lattices; a description of this exercise is given in the following section.

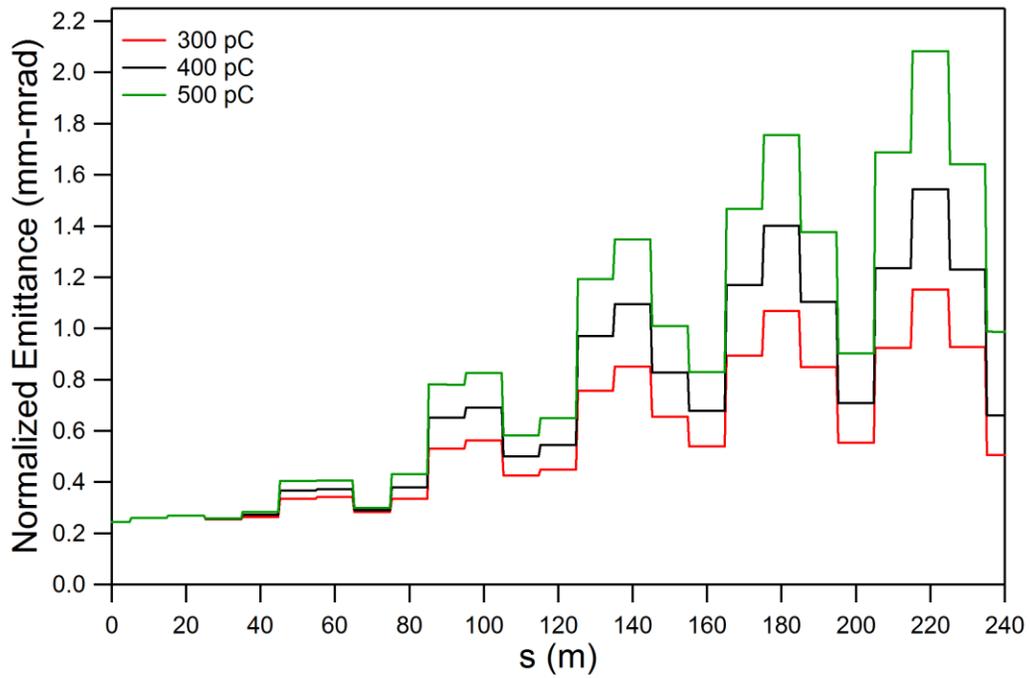

Figure 28: Horizontal normalized emittance evolution through the arc for bunch charges of 300, 400 and 500 pC.

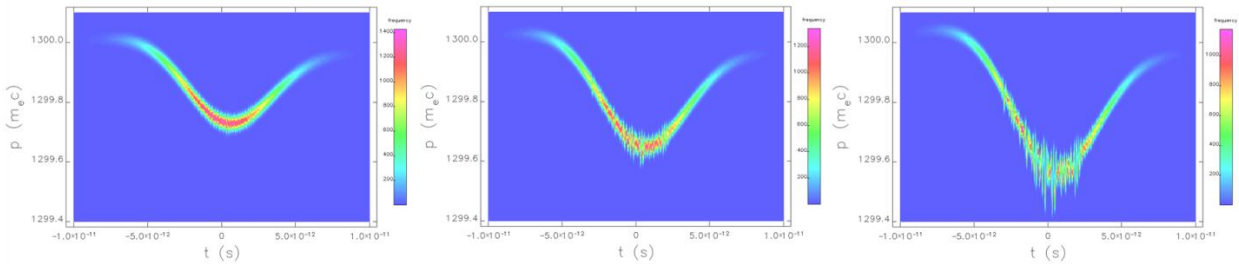

Figure 29: Longitudinal phase space at the exit of the arc for bunch charges of 300 pC (left), 400 pC (center) and 500 pC (right). All exhibit microbunching.

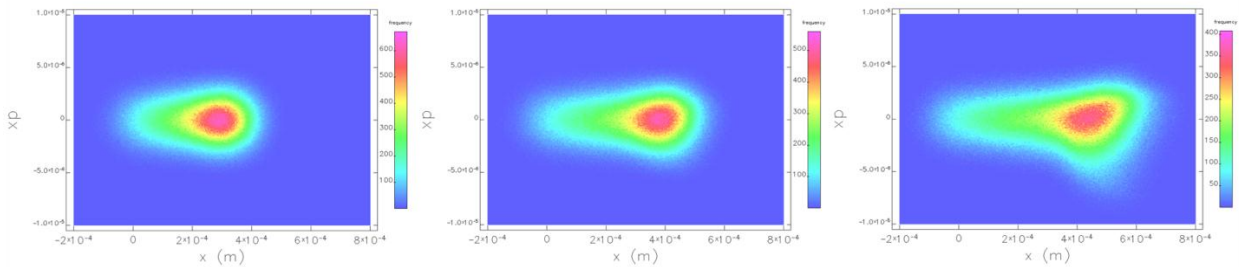

Figure 29b: Horizontal phase space at the exit of the arc for bunch charges of 300 pC (left), 400 pC (center) and 500 pC (right) corresponding to normalized emittances of 0.50, 0.66 and 0.99 mm-mrad, respectively.

**Analysis of Microbunching Gain**

It is well known that the microbunching instability can be derived from modulation in the longitudinal phase space distribution. Thus the Vlasov equation can be naturally fit to this problem by applying standard perturbation formalism. We employ the linearized Vlasov equation with 1-D CSR wake to estimate the induced microbunching gains in our demonstrated examples through some modifications and generalization of that published by Heifets et al. Based on [12], the model assumes coasting beam with Gaussian distributions over the transverse coordinates as well as over momentum deviation,

$$f_0 = \frac{n_b}{(2\pi)^2 \epsilon_{x0} \epsilon_{y0}} \exp(-\frac{x_0^2 + (\beta_{x0}\theta_{x0})^2}{2\epsilon_{x0}\beta_{x0}}) \exp(-\frac{y_0^2 + (\beta_{y0}\theta_{y0})^2}{2\epsilon_{y0}\beta_{y0}}) \rho_G(\delta + uz_0) \quad (1)$$

where

$$\rho_G(\delta) = \frac{1}{\sqrt{2\pi}\sigma_\delta} \exp(-\frac{\delta^2}{2\sigma_\delta^2})$$

To incorporate the CSR effect into the formulation, it is convenient to introduce the 1-D CSR impedance in free space [13],

$$Z(k,s) = -\frac{ik^{1/3}A}{\rho^{2/3}} \quad (2)$$

where $A \approx$ -0.94 + 1.63i and it is assumed that the radiation formation length is negligible.

As for the linearization of Vlasov equation, we consider a small sinusoidal perturbation of the equilibrium distribution

$$\rho = \rho_0 + \rho_1 = f_0 + f_k e^{ikz} \quad (3)$$

where $\rho_1 \ll \rho_0$. Thus the linear stability of the beam is determined by the solution of $\rho_1$ for the linearized Vlasov equation.

Applying the method of characteristics and after some algebraic manipulation, the linearized Vlasov equation can be solved and expressed in the general form of Volterra integral equation,

$$g_k(s) = g_k^{(0)}(s) + \int_0^s K(s,s')g_k(s')ds' \quad (4)$$

where $g_k(s)$ and $g_k^{(0)}(s)$ are the perturbed and unperturbed bunching factors at s related to $f_k$ respectively. . The kernel function $K$ has the form of

$$K(s,s') = \frac{ikr_e n_b}{\gamma}C(s)C(s')Z(kC(s'),s')R_{56}(s' \to s)$$
$$\exp\left\{-\frac{k^2\epsilon_{x0}}{2\beta_{x0}}\left(\beta_{x0}^2 R_{51}^2(s,s') + R_{52}^2(s,s')\right) - \frac{k^2\epsilon_{y0}}{2\beta_{y0}}\left(\beta_{y0}^2 R_{53}^2(s,s') + R_{54}^2(s,s')\right) - \frac{k^2\sigma_\delta^2}{2}R_{56}^2(s,s')\right\} \quad (5)$$

Note that because certain beam lines consist of both horizontal and vertical bends, we also include $R_{53}$ and $R_{54}$ effects in the kernel given by Ref. [12, 14]. The amplification factor we obtained in the simulation for the density perturbation is defined as

$$G(s) = \frac{|n_{1,k}(s,z)|}{C(s)n_{1,k}^{(0)}}$$

where $n_{1,k}$ is related to $g_k(s)$ and $n_{1,k}^{(0)}$ is given and $C(s)$ is the following compression factor.

$$n_{1,k} = C(s)g_k(s)e^{ikC(s)z}$$

Here, we also employ another approach to study the CSR effect. Motivated by [14], we express (4) in matrix form and construct iterative solutions from the lowest-order solution $g_k^{(0)}(s)$. For the first order iterative solution,

$$g_k^{(1)}(s) = g_k^{(0)}(s) + \int_0^s K(s,s')g_k^{(0)}(s')ds' \equiv (1+\mathbf{K})g_k^{(0)}(s)$$

Similarly, the second order iterative solution can be expressed as follows.

$$g_k^{(2)}(s) = g_k^{(0)}(s) + \int_0^s K(s,s')g_k^{(1)}(s')ds'$$

$$= g_k^{(0)}(s) + \int_0^s K(s,s')\left\{ g_k^{(0)}(s') + \int_0^{s'} K(s',s'')g_k^{(0)}(s'')ds'' \right\}ds'$$

$$= g_k^{(0)}(s) + \int_0^s K(s,s')g_k^{(0)}(s')ds' + \int_0^s K(s,s')\left\{ \int_0^{s'} K(s',s'')g_k^{(0)}(s'')ds'' \right\}ds'$$

$$= (1+\mathbf{K}+\mathbf{K}^2)g_k^{(0)}(s)$$

It is easily seen the two approaches are essentially equivalent assuming the infinite sum of kernel matrix converges.

$$g_k(s) = \cdots = (1+\mathbf{K}+\mathbf{K}^2+\mathbf{K}^3+\cdots)g_k^{(0)}(s) = (1-\mathbf{K})^{-1}g_k^{(0)}(s)$$

The second approach is advantageous in that it allows us to not only gain more physics insight, it also provides an alternative to verify our simulation results. Motivated by [14], the physical meaning of the single kernel function $K$ can be viewed as describing that an upstream density modulation induces energy modulation via CSR interaction, which in turn is subsequently turned into a downstream density modulation at through the transfer matrix $R_{56}(s,s')$. Furthermore, this approach can also give an implication of the so-called *staged* amplification mechanism, a topic which is under further study.

**Numerical Results**

We applied the above two numerical approaches to the analysis of the Example 2 and 3 beam lines. Table 4 lists and compares beam and lattice parameters for the two lattices with different oscillation amplitudes of $R_{56}$ and their periodicity. Both lattices consist of 24 dipoles with the same bending radius (~3.6 m). The coasting beam approximation can be valid assuming the modulation wavelength is much smaller than that of the nominal bunch length (~600 μm). The calculated $R_{56}$ as a function of $s$ is shown for Example 2 in Figure 13 and for Example 3 in Figure 24c, where it can be easily seen the oscillation amplitudes and periodicity of $R_{56}$ are quite different. For Example 3 – the case of aperiodic transport arc with large oscillation of $R_{56}$ – we found that the CSR-induced microbunching gain is much larger than that of the proposed periodic transport arc with small oscillation of $R_{56}$. See Figure 30. Furthermore, the amplification gain is expected to possibly increase in dipoles and kept constant elsewhere since we have ignored CSR-drift effect. To look over the possibly covered spectrum for inducing CSR microbunching, we consider the wavelength range from 1 to 200 μm and locate $G(s)$ at the exit of the transport arcs to obtain the gain spectrum $G(\lambda)$ (Figure 31) It can be seen that, for smaller modulation wavelengths, the Landau damping induced from either finite emittance or energy spread can be enhanced so that the overall amplification gain drops. While for larger modulation wavelengths the overall gain still drops due to negligible CSR interaction. One interesting feature of the microbunching amplification in the two cases are that it takes six stages of iteration of the integral equation for the gain to reach the self-consistent solution of the Eq. (4). This may indicate coupling effect among multiple dipoles.

Table 4: Beam and lattice parameters for the high energy recirculation arcs; v-1 for aperiodic lattice with large oscillation of $R_{56}$, v-2 for periodic lattice with small oscillation of $R_{56}$.

| | Example 2 | Example 3 |
|---|---|---|
| beam energy (GeV) | 1.3 | 1.3 |
| FWHM bunch current (A) | 71 | 71 |
| line density (cm$^{-1}$) | $1.4766 \times 10^{10}$ | $1.4766 \times 10^{10}$ |
| normalized emittance (mm-mrad) | 0.25 | 0.25 |
| initial beta function $\beta_{x0}$ (m) | 65.0 | 35.81 |
| initial alpha function $\alpha_{x0}$ | 0 | 0 |
| uncorrelated relative energy spread | $8.97 \times 10^{-6}$ | $8.97 \times 10^{-6}$ |
| arc structure | 2$^{nd}$ order achromat, periodically isochronous and achromatic; achromatic/isochronous superperiods | 2$^{nd}$ order achromat, globally isochronous and achromatic; dispersive/nonisochronous superperiods |
| RMS bunch length (mm) | 0.9 | 0.9 |
| number of dipoles | 24 | 24 |

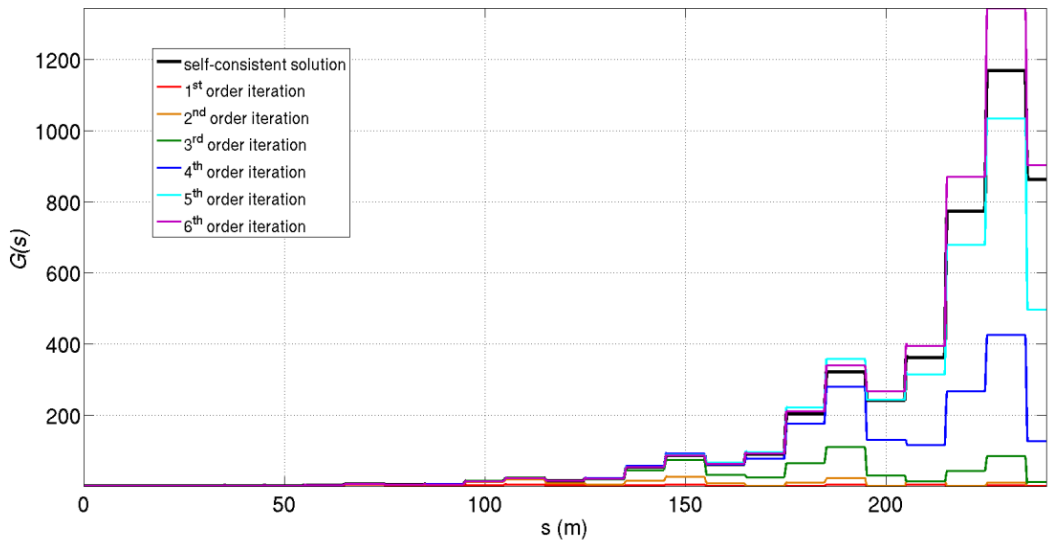

Figure 30a: Gain function G(s) as a function of s for Example 3, the high dispersion lattice with large $R_{56}$, $\sigma_\delta = 9 \times 10^{-6}$, $\lambda_m = 35$ μm.

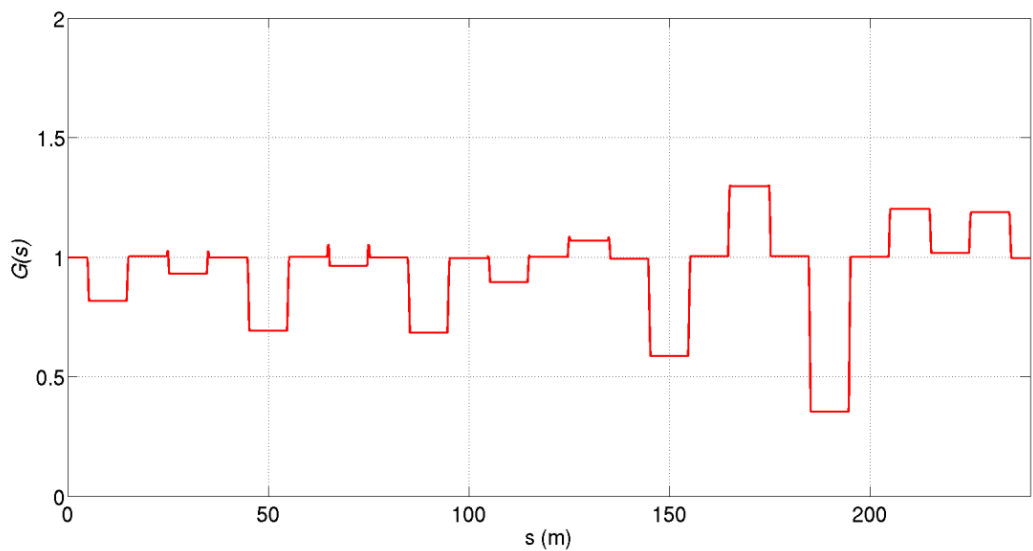

Figure 30b: Gain function G(s) as a function of s for Example 2, the periodically isochronous lattice with small $R_{56}$, $\sigma_\delta = 9 \times 10^{-6}$, $\lambda_m = 35$ μm.

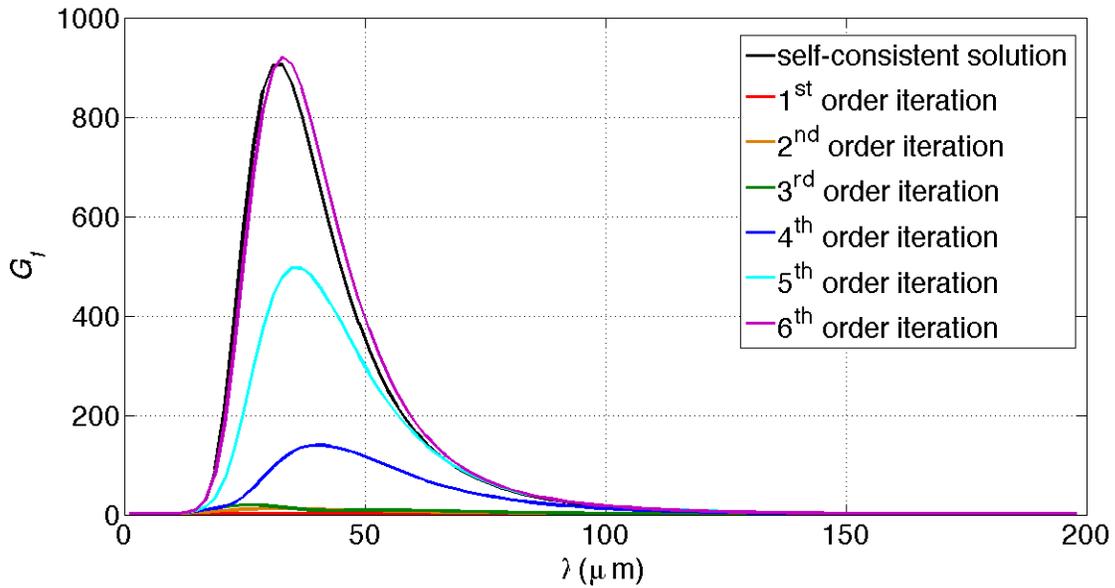

Figure 31: Gain spectrum, $G(\lambda)$ as a function of modulation wavelength for Example 3 − high dispersion lattice with large $R_{56}$, $\sigma_\delta = 9 \times 10^{-6}$.

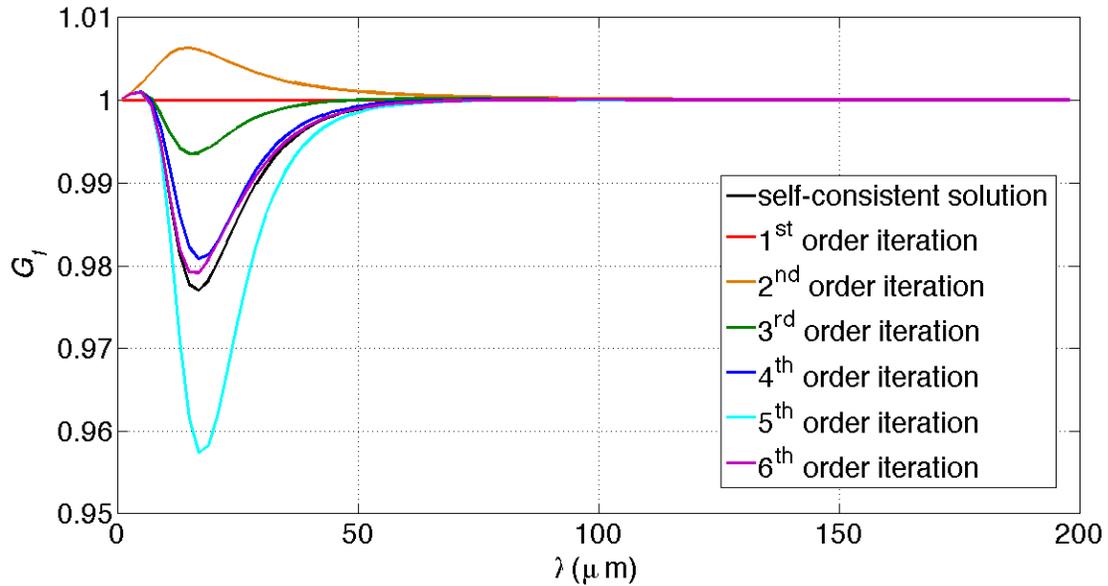

Figure 31: Gain spectrum, $G(\lambda)$ as a function of modulation wavelength for Example 2 − periodically isochronous lattice with small $R_{56}$, $\sigma_\delta = 9 \times 10^{-6}$.

**Conclusions**

Second order achromats composed of individually linearly isochronous and achromatic superperiods provide excellent suppression of CSR-induced emittance growth and appear to limit the gain of the microbunching instability, and thus allow the recirculation of high-brightness electron beams. Results may be extrapolated to apply to systems such as electron-ion colliders and free-electron laser drivers.

## Acknowledgements

We thank Dr. Hugh Montgomery for his ongoing encouragement to document this work, Dr. Roy Whitney for useful discussions, and Dr. Fay Hannon for a skillful rendering of some of the figures. This manuscript has been authored by Jefferson Science Associates, LLC under Contract No. DE-AC05-06OR23177 with the U.S. Department of Energy. The United States Government retains and the publisher, by accepting the article for publication, acknowledges that the United States Government retains a non- exclusive, paid-up, irrevocable, world-wide license to publish or reproduce the published form of this manuscript, or allow others to do so, for United States Government purposes.